# Quantum-trajectory analysis for charge transfer in solid materials induced by strong laser fields

Shicheng Jiang,[1] Chao Yu,[1] Guanglu Yuan,[1] Tong Wu,[1] Ziwen Wang,[1] and Ruifeng Lu[1,2,*]

[1]*Department of Applied Physics, Nanjing University of Science and Technology, Nanjing 210094, P R China*

[2]*State Key Laboratory of Molecular Reaction Dynamics, Dalian Institute of Chemical Physics, Chinese Academy of Sciences, Dalian 116023, P R China*

## Abstract

We investigate the dependence of charge transfer on the intensity of driving laser field when $SiO_2$ crystal is irradiated by an 800 nm laser. It is surprising that the direction of charge transfer undergoes a sudden reversal when the driving laser intensity exceeds critical values with different carrier envelope phases. By applying quantum-trajectory analysis, we find that the Bloch oscillation plays an important role in charge transfer in solid. Also, we study the interaction of strong laser with gallium nitride (GaN) that is widely used in optoelectronics. A pump-probe scheme is applied to control the quantum trajectories of the electrons in the conduction band. The signal of charge transfer is controlled successfully by means of theoretically proposed approach.

---

[*] rflu@njust.edu.cn

# 1. INTRODUCTION

It is a common sense that the minimum time window required to change the electronic signal determines the rate of information exchange. The metal-oxide-semiconductor field-effect transistors (MOSFETs) are the basic devices in modern telecommunication technology [1] .The cutoff-frequencies of the MOSFETs can be higher than 100 GHz in theory [2–4]. While, the maximum speed of the processor has been limited to ~3 GHz for a long time because of two main factors: the charging time of the interconnect wires and heat dissipation [5]. Fortunately, laser technology provides an alternative strategy to control the electric signals. In 2011, Ghimire *et al.* first observed non-perturbative high harmonics in ZnO crystal [6]. In the later years, the spectral plateau has been extended to the EUV range [7,8]. These works demonstrated that it is possible to control the electric current on a subpicosecond timescale. In 2013, F. Krausz and his team revealed an ultrafast 'turning on' and 'turning off' of the current [9,10]. The conductivity and, consequently, current can be switched on and off in a dielectric using optical fields on a timescale of, the order of or less than 1 fs. Since the current signals are switched so fast that the current measurement techniques don't allow for detecting such currents in real time. The transferred charges induced by driving laser become the signal used to investigate the response of the materials to the driving laser. It has been demonstrated in experiment that there is charge transfer along the direction of the driving few-cycle laser [9]. Besides, the signals of charge transfer show oscillation behavior with changing of carrier-envelope phase (CEP) of the laser [9,11]. These characteristics won't be changed for different materials [12]. The CEP-dependent charge transfer can also be used to detect the CEP of few-cycle laser pulses under ambient conditions [11]. In theory [13], the carrier-envelope phase control of the electric current was interpreted as a result of quantum-mechanical interference of multiphoton excitation channels. For such important potential applications, laser-induced charge transfer has attracted much attention in recent years [9–14].

In this work, the laser induced charge transfer is investigated theoretically by solving semiconductor Bloch equations (SBEs) [15–20]. In our previous study, the shapes of $k$-space-dependent dipole moments have been demonstrated to play an important role in harmonic generation [21]. It is necessary to perform calculations with accurate band structure and $k$-space-dependent dipole moments. Here after, all the band structures and $k$-dependent dipole moments of the target materials were obtained from high-level first-principles calculations using the Vienna ab initio simulation package (VASP) [22–24]. Firstly, α-quartz $SiO_2$ which was widely used in recent experiments is taken as a target material [7,9–11]. As the intensity of the driving field is tuned from $1\times10^{13}$ W/cm$^2$ to $6\times10^{13}$ W/cm$^2$, the direction of the charge transfer undergoes a sudden reversal when the driving laser intensity continues increasing. Then, a method which we call quantum-trajectory analysis is employed to seek the mechanism underlying this phenomenon. We find that the quantum-trajectories of the electron wavepackets in conduction band can be controlled by the driving field. Moreover, a pump-probe scheme is applied to control the quantum trajectories of the excited electrons. For large bandgap materials, the laser induced currents will be controlled at the expense of very high intensity fields, thus we subsequently focus on the target material, gallium nitride (GaN), which is widely used in optoelectronics. As a result, the signal of charge transfer is successfully controlled.

## 2. THEORETICAL MODEL

The first target we studied is $SiO_2$ crystal in α-quartz structure with P3121 symmetry group. Since the driving laser in our work is linearly polarized, a one-dimension calculation is sufficient. Here, the direction Γ-M in reciprocal space is chosen. The two-band SBEs describing the laser-crystal interaction read

$$i\hbar\frac{\partial}{\partial t}p_k = \left(\varepsilon_c(k) - \varepsilon_v(k) - i\frac{1}{T_2}\right)p_k - \hbar\left(1 - f_k^e - f_k^h\right)d(k)E(t) + iE(t)\nabla_k p_k, \tag{1}$$

$$\hbar\frac{\partial}{\partial} f_k^{e(h)} = -2\,\text{Im}\left[\hbar d(k)E(t)p_k^*\right] + E(t)\nabla_k f_k^{e(h)}, \tag{2}$$

where $f_k^e$ ($f_k^h$) is the occupation of electron (hole) in the lowest conduction band (highest valance band). $P_k$ is the microscopic polarization between conduction and valance band. $\varepsilon_c(k)$ and $\varepsilon_v(k)$ are the **k**-dependent energy bands of conduction band and valance band, respectively. $d(k)$ is **k**-dependent dipole moment.

The energy bands and dipole moments are calculated by VASP code. The band structures and dipole moments of $SiO_2$ can be found in our previous work [21]. Once the electrons in valance band are excited to conduction band, the electrons in conduction band and holes in valance band will be driven by the laser. The total currents induced by the driving field contains three parts: intraband electric currents in conduction band $j_c(t)$ and valance band $j_v(t)$, and $j_{cv}(t)$ induced by macroscopic interband polarization. These three items are given by

$$j_{c(v)}(t) = \int_{BZ} e\frac{\partial \varepsilon_{c(v)}(k)}{\partial k} f_k^{e(h)}(t)dk, \tag{3}$$

$$j_{cv}(t) = \int_{BZ}\left[d(k)\frac{\partial p_k(t)}{\partial t} + c.c.\right]dk, \tag{4}$$

where $e = +1$ for valance band and $e = -1$ for conduction band. The transferred charge induced by driving laser is

$$Q(t) = \int_{-\infty}^{t}\left(j_c(t') + j_v(t') + j_{cv}(t')\right)dt'. \tag{5}$$

The component of $j_v(t)$ is negligible for the low mobility of holes in the valance band. Besides, $j_{cv}$ mainly describes the macroscopic polarization response which will be vanished when the laser is over [25]. So, we restrict our analysis to the transferred charge induced by $j_c(t)$ in this work. As far as we know, the signals of charge transfer calculated by solving SEBs have not been compared with available experimental data. To provide more evidences for that the SEBs can be used to calculate the signal of charge transfer, we compare our theoretical results with the first experimental data [9] in Appendix A.

## 3. RESULTS AND DISSCUSSION

In figure 1(a), the dependence of transferred charge on the intensity of laser with CEP = 0.4π is displayed. The black squares represent the values of transferred charge. The red dots represent the values of $C\int_{-\infty}^{t} A^3(t')dt'$, where $A(t)$ is the vector potential of the driving laser, and $C$ is an adjustable parameter. As already demonstrated [26], the charge can be obtained by $Q(t) \sim \int_{-\infty}^{t} A^3(t')dt' + ...$ when the laser intensity is weak. Whereas we observe that as the intensity of laser is increased, the direction of the charge transfer undergoes a sudden reversal, and the charge transfer won't follow $\int_{-\infty}^{t} A^3(t')dt'$ any more. The critical intensity can be seen to be 4.4×10$^{13}$ W/cm$^2$ (1.84 V/Å) for CEP = 0.4π. In figure 1(b), we also present the dependence of transferred charge in SiO$_2$ on CEP for laser intensities of 3.6×10$^{13}$ W/cm$^2$, 4.0×10$^{13}$ W/cm$^2$, 4.8×10$^{13}$ W/cm$^2$, and 5.0×10$^{13}$ W/cm$^2$, respectively. It is clear that the transferred charge is periodically and nearly synchronously (with little shift) varied with CEP. We note that Wachter and his co-workers also found a sudden reversal of the directional currents above critical threshold intensity in their simulation based on time-dependent density functional theory [27]. In their case, CEP effect was removed by applying a longer laser pulse. They attributed the reversal to tunnelling excitation which is sensitive to the relative orientation between laser polarization and chemical bonds.

To investigate the charge transfer reversal in detail, a quantum-trajectory analysis is applied. As we know, once the electron is excited into the conduction band, it will be driven by the laser field. Figures. 2(a) and 2(b) present the laser-driven motion of the electron density in the conduction band for laser intensities of $3.6 \times 10^{13}$ W/cm$^2$ and $4.8 \times 10^{13}$ W/cm$^2$, respectively. It is found that there exist three main quantum trajectories labeled by trajectory **1, 2** and **3**. Like the "long or short quantum trajectory" of HHG from atoms and molecules, the electrons "born" in the conduction band at different times will experience different trajectories, leading to different contribution to the charge transfer. The black lines in figures 2(a) and (b) represent the "semiclassical trajectories" born at different times. The "semiclassical trajectories" can be expressed as following:

$$k(t) = k_0 - \int_{t_b}^{t} E(t)dt, \tag{6}$$

where $k_0$ is the initial position of the excited electron, $t_b$ is the time when the electron is excited from valance band to conduction band. In our calculations, $k_0$ is set to be −0.045, and $t_b$ is set to be 6.44 fs, 6.65 fs and 6.83 fs for these three trajectories, respectively. By comparing figure 2(a) and figure 2(b), we can see that as the intensity of the laser increases, the quantum trajectories will be changed, especially for the trajectory **1**. This phenomenon will shed light on the charge transfer reversal. In a semiclassical picture, the displacement of the electron in the conduction band driven by the laser can be expressed as:

$$D = \int_{t_b}^{t_{max}} v_c(k(t))dt, \tag{7}$$

where $v_c = \frac{\partial \varepsilon_c(k)}{\partial k}$. The positive/negative $D$ will contribute to the negative/positive transferred charge. We then investigate the dependence of $D$ on laser intensities for these three semiclassical trajectories, as presented in figure 3. It is clearly seen that the semiclassical displacement for trajectory **1** changes from positive values to negative values as the intensity of the laser is increased. As a result, the whole transferred charge will be changed from negative to positive as the laser intensity reaches a threshold value.

The quantum-trajectory analysis has been applied to explain the reversal of the charge transfer successfully above. Now, we know that the signal of charge transfer is determined by the quantum trajectory of the electronic wave packets on the conduction band. Based on the analysis above, two important aspects should be kept in mind. First, the electrons excited from valance band to conduction band at different times would travel different quantum trajectories. Second, all these quantum trajectories would contribute to the charge transfer signal. A pump-probe scheme can be utilized to control the signal of charge transfer. A 1.68 fs/337 nm ultraviolet pulse is used to pump the electrons from valance band to conduction band. Then, a 15 fs/2500 nm mid-infrared pulse is used to drive the electrons in the conduction band. As the delay time between these two pulses is altered, the quantum trajectory of the electron on the conduction band will be controlled. The sketch of the pump-probe scheme is presented in figure 4(a). To be honest, efficient interband tunneling from valance band to conduction band requires very high intensity of pump laser for materials with large band gap. So we turn to GaN crystal which is widely used in optoelectronics with the calculated band gap $E_g \approx 3.4$ eV in agreement with previous work [28]. Very recently, GaN was also taken as sample to investigate the control of current in semiconductor by the CEP of the laser [29].

The accurate band structure and *k*-space-dependent dipole moments of GaN obtained from high-level first-principles calculations can be found in the Appendix B. The intensities of the ultraviolet and mid-infrared pulses are set to be $5\times10^{11}$ W/cm$^2$ and $5\times10^{10}$ W/cm$^2$, respectively, so that the ultraviolet pulse could pump the electron efficiently, but due to its short wavelength, it has little influence on the quantum trajectories when they are driven by the mid-infrared pulse. The mid-infrared pulse will hardly contribute to the excitation process for its low intensity and long wavelength. The black squares in figure 4(b) shows the dependence of the charge transfer on the delay time. The charge transfer is controlled successfully by altering the time delay between the ultraviolet pulse and mid-infrared pulse. The blue line in figure 4(b) is the vector potential of the mid-infrared laser with CEP = 0. We find that the charge transfer highly depends on the value of the vector potential of the laser when the electrons are excited into conduction band. Next, we will explain this finding. Assuming that the charge transfer is mostly induced by the intraband moving of the electrons on the conduction band, the transferred charge $Q$ will read as:

$$Q=\int_{t_b}^{t_{max}} \left(-v_c(k(t))f_0\right)dt, \tag{8}$$

where $v_c(k) = \dfrac{\partial \varepsilon_c(k)}{\partial k}$, $f_0$ is the density of the electrons excited to the conduction band by ultraviolet pulse. The excitation is assumed to occur at time $t_b$, and $f_0$ won't change in the following time. Since the energy bands can be expressed as Eq. B1 (Appendix B), the odd function $v_c(k)$ can be expanded as:

$$v_c(k) = \alpha_1 k + \alpha_3 k^3 + \alpha_5 k^5 + \cdots + \alpha_{2n+1} k^{2n+1}, \tag{9}$$

where $\alpha_{2n+1} = \dfrac{1}{(2n+1)!} \dfrac{\partial^{2n+1} v(k)}{\partial k}\bigg|_{k=0}$. So Eq. (8) is changed to be

$$Q = f_0 \int_{t_b}^{t_{max}} \left( \alpha_1 k(t) + \alpha_3 k^3(t) + \alpha_5 k^5(t) + \cdots + \alpha_{2n+1} k^{2n+1}(t) \right) dt. \tag{10}$$

Inserting Eq. (6) into Eq. (10), we can get

$$Q \approx f_0 \int_{t_b}^{t_{max}} \left( \alpha_1(-A(t) + A(t_b)) + \alpha_3(-A(t) + A(t_b))^3 + \cdots + \alpha_{2n+1}(-A(t) + A(t_b))^{2n+1} \right) dt, \tag{11}$$

where $A(t)$ is vector potential of the mid-infrared laser, and $k_0$ is ignored. If we just keep the linear part,

$$Q \approx f_0 \int_{t_b}^{t_{max}} \alpha_1(-A(t) + A(t_b)) dt = -\alpha_1 f_0 \int_{t_b}^{t_{max}} A(t) dt + \alpha_1 f_0 (A(t_b) t_{max} - A(t_b) t_b). \tag{12}$$

The first part in the right side of Eq. (12) will contribute little to $Q$. So Thus, we can understand that the charge transfer will highly depend on $A(t_b)$, which is the value of vector potential of the mid-infrared laser when the electrons are excited into conduction band. Of course, charge transfer will also depend on $t_b$ which is the time when the electrons are excited from valance band to conduction band. For example, although the absolute values of the vector potential marked by red circles are the same in figure 4(b), $|Q|$ for S1 is slightly larger than that for S2.

## 4. SUMMARY

In conclusion, the SBEs have been solved to calculate the laser induced charge transfer. In general, the theoretical results match the early experimental data very well, so it is appropriate by using this approach to investigate the phenomenon of laser induced charge transfer in semiconductors. The direction of laser induced charge transfer will undergo a sudden reversal when the driving laser intensity exceeds critical values in $SiO_2$ crystal irradiated by 800 nm laser. By applying quantum-trajectory analysis, we address that the electrons excited from valance band to conduction band at different times would travel different quantum trajectories. All these quantum trajectories would contribute to the charge transfer signal. As laser intensity is increased, one of these trajectories will travel beyond the first Brillouin zone, and Bloch oscillation occurs. Consequently, the signal of charge transfer will reverse its direction. Further, a pump-probe scheme is proposed to control the signal of charge transfer in GaN semiconductor. As the delayed time between these two pulses is altered, the quantum trajectories of the electrons on the conduction band will be tuned; hence the signal of charge transfer is controlled successfully. What's more, the charge transfer is found to highly depend on $A(t_b)$ which is the value of vector potential of the mid-infrared laser when the electrons are excited into conduction band. We expect this theoretical work would provide useful insights into potential utilization of optical field controlled ultrafast electronics using dielectrics or semiconductors.

**ACKNOWLEDGMENTS**

This work was supported by NSF of China Grant No. 21373113, the Fundamental Research Funds for the Central Universities (No. 30920140111008 and 30916011105).

**APPENDIX A: THEORETICAL RESULTS VS. EXPERIMENTAL DATA**

In this part, the results calculated by solving SBEs are compared with the first experimental data for SiO$_2$ crystal [9]. The parameters of the laser are set to be the same as in the experiment. A 4 fs/780 nm laser pulse is used. The dependence of the maximum amplitude of the transferred charge on the laser intensity is presented in figure 5(a) in red stars. What we should emphasize is that all the calculated results are multiplied by 3500 to match the experimental data. Since the effective surface area is difficult to determine experimentally, it is reasonable that the theoretical results are multiplied by a scale factor for comparing the experimental data. In figure 5(b), CEP dependence of charge transfer is shown when the laser intensity is about 1.7 V/Å. In order to simulate the experimental environment as much as possible, the screening field induced laser is taken into account [30]. While in the main text, the screening field is not taken into account for simplicity. In general, our calculated results by solving SBEs match the experimental data very well.

**APPENDIX B: BAND STRUCTURE AND DIPOLE MOMENTS FOR GaN**

Only the lowest conduction band and the highest valance band are considered in calculations. The polarization of the laser is along Γ-A. Therefore, we merely present the dispersions of these two bands for GaN in figure 5. The valance and conduction bands are indicated by black squares and red dots, respectively. The blue line is the dipole moments between the highest valance band and the lowest conduction band. The *k*-dependent band structures can be expanded as:

$$E_{c(v)}(k) = \sum_{j=0}^{+\infty} a_{c(v)}(i) \cos\big((i-1) \cdot k \cdot b\big), \tag{B1}$$

where *b* (5.166 Å) is the optimized lattice constant along Γ-A that is close to reported value [28], $a_{c(v)}(i)$ are expansion coefficients. These coefficients can be found in Table I.

**REFERENCES**

[1] Kahng D 1963 *United States Patent 3* **102** 230

[2] Taur Y and Ning T H 1998 *Fundamentals of Modern VLSI Devices* (Cambridge Univ. Press)

[3] Liou J J and Schwierz F 2003 *Modern Microwave Transistors: Theory, Design and Performance* (Wiley)

[4] Kim D H and del Alamo J A 2010 *Device Lett.* **31** 806

[5] Krausz F and Stockman M I 2014 *Nat. Photonics* **8** 205

[6] Ghimire S *et al* 2011 *Nat. Phys.* **7** 138

[7] Luu T T *et al* 2015 *Nature* **521** 498

[8] Ghimire S *et al* 2014 *J. Phys. B: At., Mol. Opt. Phys.* **47** 204030

TABLE I. Expansion coefficients for fitting the lowest conduction band (CB) and the highest valance band (VB) of GaN.

| CB | $a_c(0)$ | $a_c(1)$ | $a_c(2)$ | $a_c(3)$ | $a_c(4)$ | $a_c(5)$ | $a_c(6)$ | $a_c(7)$ | $a_c(8)$ |
|---|---|---|---|---|---|---|---|---|---|
|  | 4.43 | −1.173 | 0.0014 | −0.047 | 0.0137 | −0.0127 | 0.008 | −0.0065 | 0.0049 |
| VB | $a_v(0)$ | $a_v(1)$ | $a_v(2)$ | $a_v(3)$ | $a_v(4)$ | $a_v(5)$ | $a_v(6)$ | $a_v(7)$ | $a_v(8)$ |
|  | −0.287 | 0.251 | −0.047 | 0.021 | −0.012 | 0.0077 | −0.0057 | 0.0045 | −0.0038 |

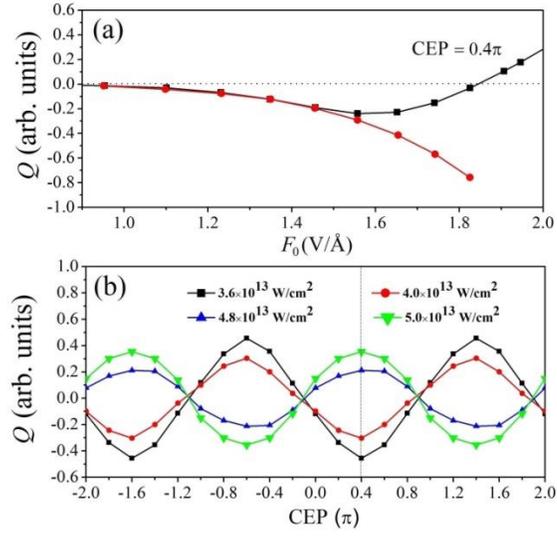

Figure 1. (a) Dependence of transferred charge (black squares) and $C\int_{-\infty}^{t} A^3(t')dt'$ (red dots) in SiO$_2$ on the intensities of laser fields when CEP = 0.4π. (b) Dependence of transferred charge in SiO$_2$ on CEP for laser intensities of 3.6×10$^{13}$ W/cm$^2$, 4.0×10$^{13}$ W/cm$^2$, 4.8×10$^{13}$ W/cm$^2$, and 5.0×10$^{13}$ W/cm$^2$ respectively.

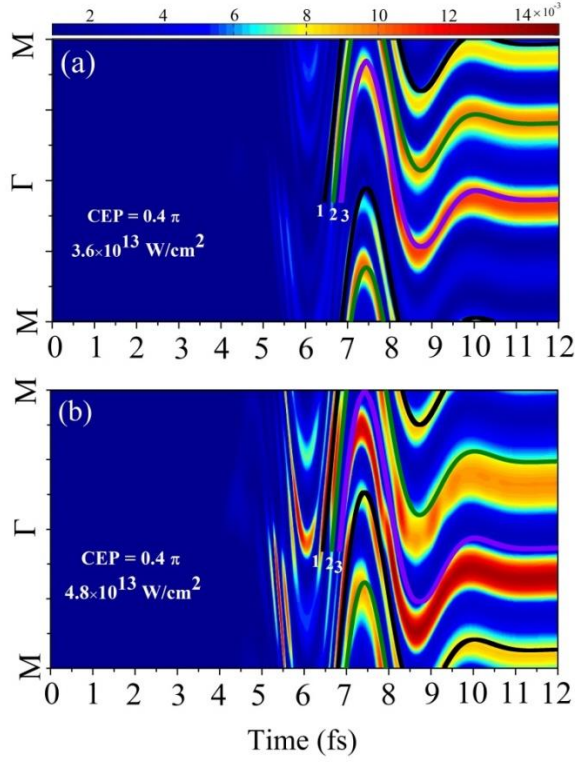

Figure. 2. Laser-driven motion of the electron density in the conduction band of SiO$_2$ for laser intensities of (a) $3.6 \times 10^{13}$ W/cm$^2$ and (b) $4.8 \times 10^{13}$ W/cm$^2$. The black lines represent semiclassical trajectories calculated by Eq. (6) in the main text. There exist three main quantum trajectories labeled by **1, 2** and **3.**

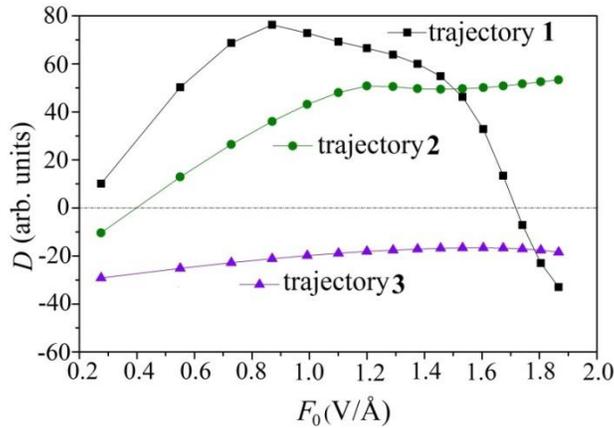

Figure. 3. Dependence of the semiclassical displacement $D$ in SiO$_2$ calculated by Eq. (7) in the main text on laser intensities for these three semiclassical trajectories.

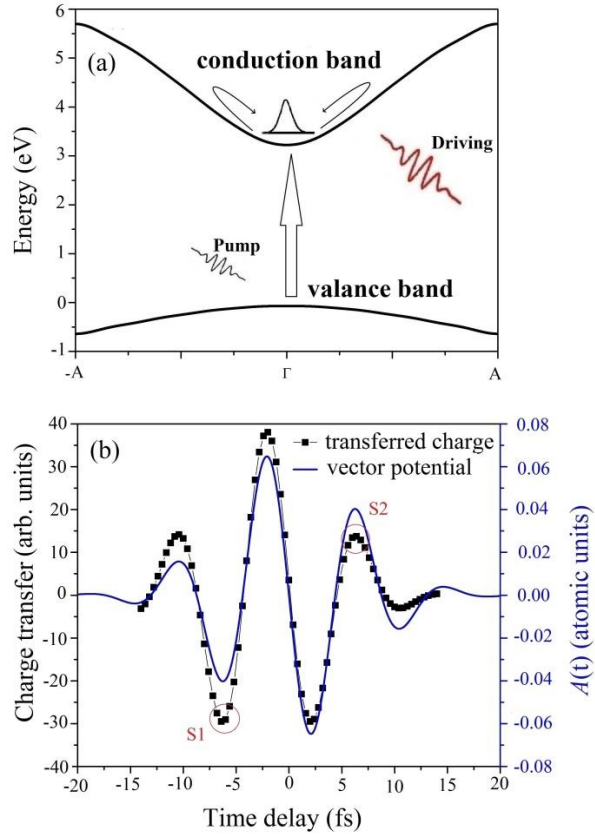

Figure. 4. (a) The sketch of pump-probe scheme for GaN semiconductor. A 1.68 fs/337 nm ultraviolet pulse is used to pump the electrons from valance band to conduction band. Then, a 15 fs/2500 nm mid-infrared pulse is used to drive the electrons in the conduction band. (b) The black squares represent the dependence of the charge transfer signals on the time delay between ultraviolet pulse and mid-infrared pulse. The blue solid line is the vector potential of the mid-infrared pulse.

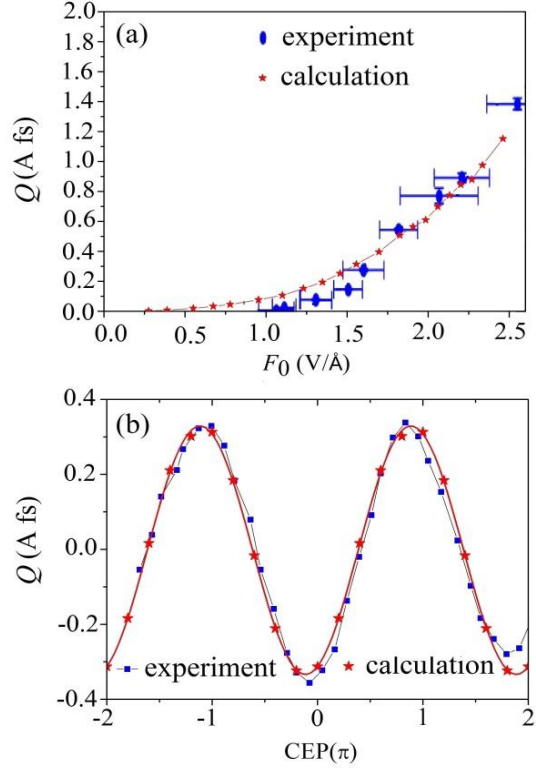

Figure. 5. (a) Laser-induced charge transfer $Q$ in $SiO_2$ as a function of laser intensity. (b) CEP dependence when the laser intensity is about 1.7 V/Å. The other parameters are the same as Ref. [9].

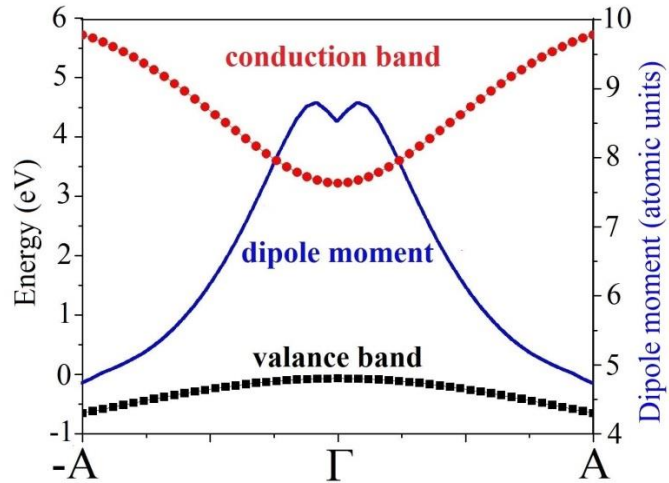

Figure. 6. The band structure of GaN semiconductor along $\Gamma$-A direction. The solid blue line shows the $k$-dependent dipole moment between the conduction band and valance band.

16